# A Three-Photon Rydberg Atom-Based Radio Frequency Sensing Scheme with Narrow Linewidth

Stephanie M. Bohaichuk[1], Fabian Ripka[1], Vijin Venu[1], Florian Christaller[1], Chang Liu[1], Matthias Schmidt[1], Harald Kübler[1,2], James P. Shaffer[1]

[1]*Quantum Valley Ideas Laboratories, 485 Wes Graham Way, Waterloo, ON N2L 6R2, Canada*
[2]*5. Physikalisches Institut, Universität Stuttgart, Pfaffenwaldring 57, 70569 Stuttgart, Germany*

**We demonstrate Rydberg atom-based radio frequency sensing with a colinear three-photon scheme in a room temperature cesium vapor cell that minimizes residual Doppler broadening of the probe laser absorption feature. A sub-200 kHz spectral linewidth is observed and extends the self-calibrated Autler-Townes sensing regime to weaker fields by a factor of ~18 compared to the theoretical limit of the most commonly used two-photon scheme. The sensitivity of the method to microsecond pulses is shown to be sufficient to detect radio frequency Rabi frequencies of $2\pi \times 0.44$ MHz using a 480 kHz bandwidth at 108.9 GHz, demonstrating the ability to sense time-dependent signals suitable for radar and communications.**

Calibration is a crucial process for a variety of technologies relying on radio frequency (RF) electromagnetic fields, including those in communications, radar, health care, remote sensing, and non-destructive testing. However, absolute RF electromagnetic field measurements have relatively low accuracy and are costly, since they have to be traced back to a standard. Novel approaches using Rydberg atoms in atomic vapors [1–4] are promising with the potential to complement, and even displace, conventional calibrated RF antennas for test and measurement across a broad frequency range spanning MHz - THz. Atom-based RF electric field (E-field) amplitude sensing can be self-calibrated because measurements can be traced to precisely determined atomic structure and fundamental constants [1,2]. These sensors are also electromagnetically transparent, portable, and robust.

In atom-based RF metrology, electromagnetically induced transparency (EIT) or electromagnetically induced absorption (EIA) is altered by an external RF E-field that is (near) resonant with atomic Rydberg transitions [2,5–8]. The smallest measurable RF E-field in the self-calibrated Autler-Townes regime is set by the spectral linewidth of the optical absorption features. Almost all Rydberg atom-based sensing to date is based on two-photon schemes. For cesium-based sensors, counter-propagating 852 nm probe and 509 nm coupling lasers are used. The spectral linewidth is limited by residual Doppler broadening, arising from the wavevector mismatch between the two lasers. For orientationally averaged transition dipole moments, the spectral resolution is limited to $2\pi \times 3.5$ MHz [9]. Optical pumping can slightly improve the spectral resolution [10]. In the amplitude regime, two-photon schemes can detect RF E-field strengths below 1 µV/cm [10,11], with a discussion of fundamental limits found in [2,12]. Weaker RF E-fields have been measured using an auxiliary RF E-field in a heterodyne measurement [13,14]. Rydberg atom-based sensors can detect modulated RF E-fields, including sub-microsecond pulses [15–20]. The sensitivity depends on the transition dipole moment and the sensing geometry.

In this work, we demonstrate a three-photon sensing scheme in a room temperature cesium vapor cell. A colinear 895 nm, 636 nm, and 2262 nm laser geometry is used to sense 108.9 GHz RF E-fields. The residual Doppler broadening is reduced, approaching a theoretical limit of $2\pi \times 51$ kHz. The wave vectors of the two longer wavelength laser fields are matched to compensate the Doppler shifts of the 636 nm laser field [9,21]. The colinear geometry is advantageous in its straightforward alignment and large interaction volume for fixed laser powers. We achieve a $2\pi \times 190$ kHz experimental spectral linewidth. We theoretically examine the practical limits to the spectral linewidth and show that it is no longer dominated by Doppler broadening. The narrow spectral linewidth significantly extends the self-calibrated Autler-Townes regime, allowing us to resolve spectral splittings of $< 2\pi \times 400$ kHz. The improvement is nearly 20 times the orientationally averaged, theoretical limit, which is rarely achieved in Rydberg atom-based sensing experiments. We also demonstrate sensing of individual $1 - 10$ µs pulses using a matched filtering technique [20] and measure RF E-fields down to fields of ~858 µV/cm in a 480 kHz detection bandwidth. The results are superior to prior all-optical two-photon sensing schemes and comparable to current heterodyning experiments at this detection bandwidth, geometry, and RF transition dipole moment.

The experimental setup is shown in Figure 1a. Three optical excitation beams pass colinearly through a 2.5 cm cesium vapor cell. Compensation coils are placed around the vapor cell to cancel the background magnetic field, which broadens the spectral line. The 895 nm probe laser drives the $6S_{1/2}(F=4) \rightarrow 6P_{1/2}(F=3)$ transition, the 636 nm intermediate coupling beam drives the $6P_{1/2}(F=3) \rightarrow$

$9S_{1/2}(F=4)$ transition, and the 2262 nm Rydberg coupling beam drives the $9S_{1/2}(F=4) \to 42P_{3/2}$ transition. The 636 nm beam counter-propagates with the 895 nm and 2262 nm beams. All beams are circularly polarized. The $1/e^2$ beam diameters of the 895 nm, 636 nm, and 2262 nm lasers are 1.2 mm, 1.1 mm, and 1.3 mm, respectively, for narrow linewidth measurements. All the lasers are frequency stabilized to an ultra-low expansion cavity using the Pound-Drever-Hall technique. The spectral linewidths of the 895 nm, 636 nm, and 2262 nm lasers are estimated to be below $2\pi \times 10$ kHz, $2\pi \times 50$ kHz, and $2\pi \times 50$ kHz, respectively. Changes in the probe beam absorption are detected using an avalanche photodiode with a 10 MHz bandwidth.

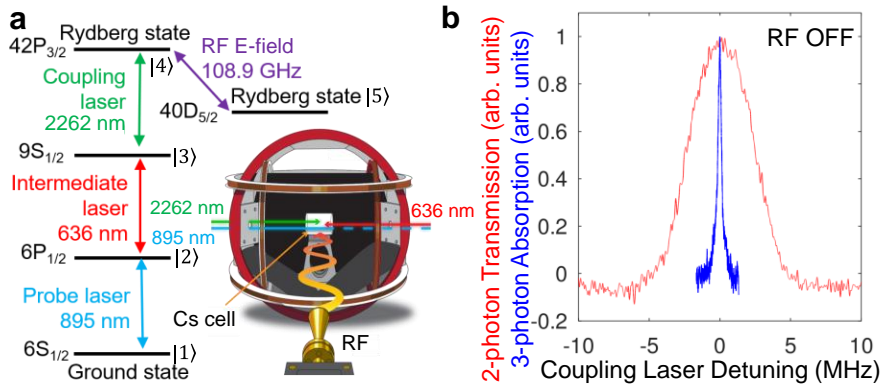

FIG. 1. (a) Energy level diagram and experimental setup. An RF E-field is emitted by a horn antenna perpendicular to the optical axis, and is sensed by monitoring the probe laser's transmission through the vapor cell. (b) The absorption feature produced by the three-photon excitation is spectrally narrower by over an order of magnitude relative to a standard two-photon 509 and 852 nm excitation scheme. Rabi frequencies of $\Omega_{895} = 2\pi \times 1.0$ MHz, $\Omega_{636} = 2\pi \times 6.7$ MHz, and $\Omega_{2262} = 2\pi \times 0.04$ MHz are used for the three-photon excitation, while $\Omega_{852} = 2\pi \times 1.7$ MHz and $\Omega_{509} = 2\pi \times 2.1$ MHz are used for the two-photon excitation.

A 108.9 GHz RF E-field is used to resonantly drive the $42P_{3/2} \to 40D_{5/2}$ atomic transition. The RF E-field strength is adjusted by a variable attenuator and radiated by a microwave horn antenna placed $\sim 15$ cm from the vapor cell. The RF E-field propagates perpendicular to the laser beams with the electric field vertically polarized. For this Rydberg transition, the radial dipole matrix element is $632ea_0$ and we use the orientationally averaged dipole moment $\mu = 400ea_0$ for estimating the RF E-field.

A typical spectral line shape for the probe absorption for the three-photon readout is shown in Figure 1b. The 2262 nm laser is modulated at 61 kHz using an acousto-optic modulator (AOM). The probe laser transmission is processed with a lock-in amplifier. The three-photon excitation creates EIA, which exhibits an approximately Lorentzian shape with a $2\pi \times 324$ kHz full-width at half-maximum (FWHM). For comparison, the transmission line shape obtained in a standard two-photon setup for cesium using counterpropagating 509 nm coupling and 852 nm probe lasers is also shown. Under narrow linewidth conditions, the two-photon system shows a Gaussian shape with FWHM of $\sim 2\pi \times 5$ MHz, over $15\times$ larger than the three-photon system shown in Figure 1c. The fact that the spectral lineshape for the three-photon readout is Lorentzian demonstrates that it is no longer dominated by the Doppler effect.

Figure 2a shows the shape and linewidth of the three-photon signal at different Rabi frequencies, $\Omega_{636}$. As $\Omega_{636}$ decreases, the signal amplitude decreases and the linewidth reduces to $2\pi \times 190$ kHz at $\Omega_{636} = 2\pi \times 1.7$ MHz, which is 18 times smaller than the orientationally averaged residual Doppler limit of a two-photon system.

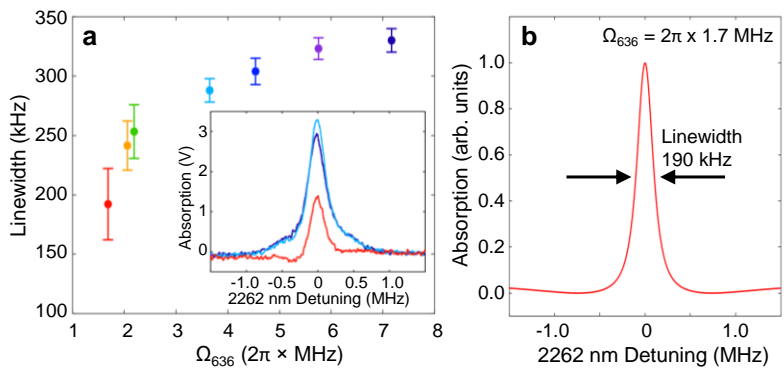

FIG. 2. (a) Extracted Lorentzian linewidth (FWHM) of the three-photon signal with no RF E-field present, measured at various intermediate coupling Rabi frequencies $\Omega_{636}$. Error bars represent one standard deviation in the distribution of linewidths from individual traces. The inset shows averaged experimental spectral line shapes. (b) Probe absorption of the three-photon system simulated using a density matrix model, producing a theoretical Lorentzian linewidth of 190 kHz. $\Omega_{895} = 2\pi \times 1.0$ MHz and $\Omega_{2262} = 2\pi \times 0.04$ MHz.

To investigate the theoretical spectral linewidth, we performed density matrix calculations shown in Figure 2b. The master equation is solved for the steady-state density matrix in the five-level ladder system in Figure 1a and then thermally averaged (see Supplement [22]). We include radiative and blackbody decay rates in the Lindblad operator for each state: $\Gamma_{21} = 2\pi \times 4.5$ MHz, $\Gamma_{32} = 2\pi \times 200$ kHz, and $\Gamma_{42} = 2\pi \times 2$ kHz [23]. We include a $2\pi \times 110$ kHz transit time broadening. Dephasing terms for each transition are also included based on the sum of the spectral linewidths of the lasers that interact with each level, $\Gamma_{11} = 2\pi \times 10$ kHz, $\Gamma_{22} = 2\pi \times 60$ kHz, $\Gamma_{33} = 2\pi \times 100$ kHz, and $\Gamma_{44} = 2\pi \times 50$ kHz.

In our model, we find that the spectral line shape of the probe absorption is Lorentzian with a FWHM of $\sim 2\pi \times 190$ kHz at $\Omega_{895} = 2\pi \times 1.0$ MHz, $\Omega_{636} = 2\pi \times 1.7$ MHz, and $\Omega_{2262} = 2\pi \times 0.04$ MHz, similar to the experimental linewidth under these conditions. Residual Doppler broadening is no longer the limiting factor in the spectral resolution of the Autler-Townes regime, but is one of several contributions to the linewidth, including laser linewidths, Rydberg state lifetimes and transit time broadening. The fact that the observed EIA line shape is predominantly Lorentzian rather than Gaussian demonstrates this change. Under our experimental conditions the model predicts a low Rydberg state population of $\rho_{44} \sim 3 \times 10^{-5}$ and we do not expect collisions to play a significant role (see Supplemental) [22]. When the laser linewidths are smaller, <1 kHz, transit time broadening and residual Doppler effects dominate and the three-photon linewidth reduces to 145 kHz. The spectral linewidth is ultimately limited by Rydberg state decay rates and the residual Doppler broadening.

Figure 3 shows experimental probe absorption with a 108.9 GHz continuous wave RF E-field applied. RF E-field amplitude $E$ is determined directly from the self-calibrated atomic response by fitting a pair of Lorentzians to the absorption peaks and extracting the frequency splitting $\Delta\nu$. This is converted to an RF E-field amplitude using $E = 2\pi\hbar\Delta\nu/\mu$. As progressively weaker RF E-fields are applied, the two peaks begin to overlap until they become indistinguishable at a splitting of $\sim 2\pi \times 380$ kHz, corresponding to a field of $\sim 745$ µV/cm. The data in Fig. 3 demonstrates that the minimum detectable RF E-field amplitude that can be measured in the self-calibrated regime is reduced by a factor of $\sim 9$ beyond the smallest reported self-calibrated field measured in any two-photon setup, at an equivalent dipole moment and as much as $\sim 18$ fold better than schemes that don't exploit optical pumping [8].

The sensitivity of the three-photon method for measuring weak RF E-fields in the amplitude regime relies on detecting small changes in absorption. Better sensitivity is achieved with higher intermediate and Rydberg coupling Rabi frequencies than in the Autler-Townes regime. To increase the Rabi frequency of the Rydberg coupling laser, whose power was limited, we amplified it by seeding a broadband multimode laser diode [24]. We also reduced the 2262 nm laser's beam size to a diameter of $\sim 150$ µm. The 895 nm laser beam diameter was reduced to 240 µm to minimize the non-signal carrying component of the probe beam. We chose maximum 2262 nm and 636 nm laser powers, and an optimal power for the 895 nm laser, such that RF sensitivity was maximized. This choice is consistent with both experiment and modeling (see Supplement [22]).

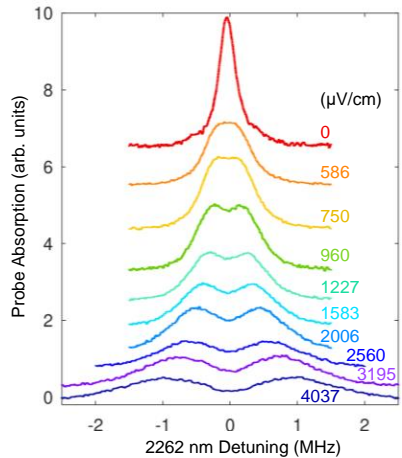

FIG. 3. Autler-Townes splitting of the probe laser absorption in the vapor cell in response to a 108.9 GHz continuous wave RF E-field whose amplitude is increased from top to bottom. The top trace is measured without an RF E-field applied, and produces a FWHM of 284 kHz. The self-calibrated RF field strength measured from the atoms via the peak splitting is labeled next to each trace. $\Omega_{895} = 2\pi \times 1.0$ MHz, $\Omega_{636} = 2\pi \times 7.2$ MHz, and $\Omega_{2262} = 2\pi \times 0.04$ MHz.

We measure the RF E-field sensitivity at $\Omega_{2262} = 2\pi \times 2.4$ MHz, $\Omega_{636} = 2\pi \times 7.2$ MHz, and $\Omega_{895} = 2\pi \times 1.9$ MHz, resulting in linewidth of $\sim 1.5$ MHz. RF E-field strength is adjusted by a linear attenuator which is calibrated at high fields where Autler-Townes splitting can be measured. Calibration is performed with the RF E-field on resonance, but a slightly higher RF E-field sensitivity is found at an RF E-field detuning of -125 kHz with the 2262 nm detuned by about half that in the opposite direction, most likely due to AC Stark shifts and slight beam misalignment. To measure the sensitivity to individual RF E-field pulses, all lasers are frequency-locked, and the RF E-field is pulse modulated on a microsecond time scale. To further increase sensitivity and stabilize the atomic density, we heat the vapor cell to 29°C. We enclose the vapor cell in a small plastic holder containing four resistive microheaters (~ 2 mm x 2 mm) in contact with the outer corners of the cell, aimed at maintaining RF transparency.

To improve the signal-to-noise ratio and obtain the sensitivity in an operational mode that is closer to the anticipated one in a fieldable device, we make use of a digital matched filter [20,25]. For a nearly square pulse, the matched filter has a well-defined detection bandwidth of $B = 1/\tau$, where $\tau$ is the pulse duration. Matched filtering relies on a known pulse shape, which we obtain from an averaged experimental atomic response, Fig. 4a inset [22]. Responses exhibit rapid ~0.5 µs rise and fall times due to the small beam size, i.e., rapid transit time. Typical output from the matched filter is shown in Figure 4a, which produces a peak at the time of maximum cross-correlation with the template. We take statistics of the peak of the matched filter in the expected arrival window, minus the average peak noise amplitude when no RF is applied, to produce Figure 4b.

We extract a sensitivity for our detection scheme by considering the weakest measurable field to be one at which the average signal height is 1σ above the filtered noise. We perform least-squares fitting of the data at weak fields in the amplitude regime, <2450 µV/cm, where the absorption change is approximately quadratic as a function of RF E-field amplitude [26]. The point at which the curve intersects the upper error bar of the noise distribution when no RF E-field is present yields a minimum detectable field for 2 µs pulses of 1225 µV/cm, or $\Omega_{RF} = 2\pi \times 0.63$ MHz. The measurement bandwidth is 500 kHz. Similarly, weaker pulses, down to 955 µV/cm, are measurable at 10 µs pulse durations (100 kHz bandwidth) due to higher total energy in the pulse, while 1754 µV/cm fields can be measured at 1.25 µs pulse durations (800 kHz bandwidth). Our overall best response to short pulsed RF E-fields was obtained at 2.1 µs pulse durations, reaching down to 858 µV/cm, or a Rabi frequency of $2\pi \times 0.44$ MHz, at a detection bandwidth of 480 kHz.

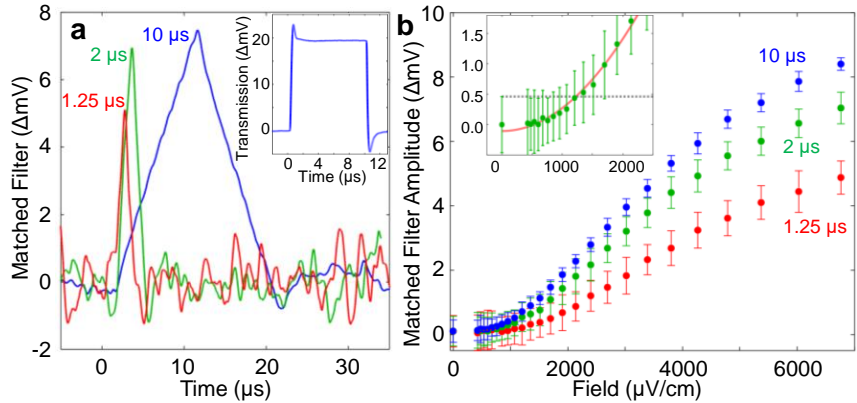

FIG. 4. (a) Matched filter output in response to single 5365 µV/cm RF pulses. The inset shows an averaged atomic response to a 3724 µV/cm RF pulse, used as a filter template. (d) Average signal height output by the matched filter in response to individual pulses at different RF E-field strengths. The error bars represent one standard deviation from 200-400 pulses. The inset shows a fit (red) to the quadratic amplitude regime at weak RF E-fields. Sensitivity is determined by the minimum field at which the average signal height exceeds a 1σ noise error bar (black dashed line in inset) at the measurement bandwidth.

The sensitivity near these measurement bandwidths corresponds to ~1.25 µV cm$^{-1}$ Hz$^{-1/2}$. Due to the nonlinear scaling of absorption with RF E-field amplitude, this value should not be extrapolated to long integration times. Evaluating the sensitivity in terms of minimum detectable RF Rabi frequency rather than RF amplitude allows for a better comparison of technical noise levels because it is independent of the transition dipole moment. For similar pulse durations, measurement bandwidths, and geometry in a two-photon setup with a matched filter, a corresponding Rabi frequency of $2\pi \times 1.37$ MHz was measured [20]. State-of-the-art heterodyne experiments have not been demonstrated at these bandwidths to our knowledge, but can reach minimum Rabi frequencies on the order of $2\pi \times 0.075$ MHz when extrapolated to a 500 kHz bandwidth [13,14]. We note that these heterodyne experiments that have reported these values use over 200 times the atoms, 200 times the probe laser power and 14 times less transit time broadening than the experiment reported here. The colinear three-photon approach can therefore offer similar or better pulse sensitivity, $2\pi \times 0.44$ MHz, in the amplitude regime than a two-photon setup, and, even given the differences in geometric conditions, is within an order of magnitude of the quoted heterodyne results.

Large improvements in RF E-field sensitivity can be expected by increasing both the 636 nm and 2262 nm Rabi frequencies, as well as by using larger beam sizes to decrease transit time broadening and increase the overall probe laser power for a fixed Rabi frequency. Larger probe laser powers reduce the photon shot noise limited lowest detectable RF E-field, while our density matrix calculations indicate that the optimum sensing point lies at larger 636 nm and 2262 nm Rabi frequencies. Using larger laser beams also increases the number of atoms and therefore signal. We estimate conservatively that gains of over 30 in sensitivity are achievable by optimizing the coupling laser Rabi frequencies and increasing the beam diameters by a factor of 10 while maintaining the probe laser Rabi frequency. A downside of increasing the laser beam size is that the reduced transit time broadening will increase the time it takes for a pulse to reach its final amplitude [20]. Given the narrow linewidth of the RF-induced changes to EIA [22], the three-photon setup is very sensitive to the frequency noise of all three lasers, since frequency noise is converted to signal amplitude noise. While our system is estimated to be within an order of magnitude of the probe laser photon shot noise limit, obtaining narrower linewidth lasers, either through laser design or improved frequency locking, is expected to allow the system to reach the probe laser shot noise limit. With these improvements, we expect the three-photon system to be comparable to or surpass current state-of-the-art heterodyne results in two-photon systems and potentially surpass the thermal noise limit.

We experimentally demonstrated a colinear three-photon scheme for Rydberg atom-based RF electrometry. The scheme minimizes wavevector mismatch between the lasers, greatly reducing residual Doppler broadening. We achieve $2\pi \times 190$ kHz EIA spectral linewidths, limited by transit time broadening, spontaneous decay, and laser linewidths. The spectral linewidth is over 18 times narrower than a standard two-photon excitation scheme, and enables measurement of weaker RF E-fields within the Autler-Townes regime. Using a matched filter tailored to the atomic response, we measured a minimum RF Rabi frequency of $2\pi \times 0.44$ MHz at a bandwidth of 480 kHz, at 108.9 GHz. At these bandwidths, the minimum Rabi frequency compares well with state-of-the art experiments when geometry and dipole moment are accounted for. We find that higher coupling laser Rabi frequencies and larger beam sizes are beneficial for achieving the highest sensitivity. The narrow linewidth extends the self-calibrated sensing regime to weaker fields, while providing high sensitivity to single RF pulses.

## ACKNOWLEDGEMENTS

This work has been supported by Defence Research and Development Canada (DRDC) under the "Innovation for Defence Excellence and Security" IDEaS program "Quantum Leap: Shrinking sensing technologies for field operation" (contract number: W7714-217517/001/SV1) and Defense Advanced Research Projects Agency (DARPA) under the "Science of Atomic Vapors for New Technologies" (SAVaNT) program (contract number: HR00112190080).

# Supplemental Material: A Three-Photon Rydberg Atom-Based Radio Frequency Sensing Scheme with Narrow Linewidth

Stephanie M. Bohaichuk[1], Fabian Ripka[1], Vijin Venu[1], Florian Christaller[1], Chang Liu[1], Matthias Schmidt[1], Harald Kübler[1,2], James P. Shaffer[1]

[1]Quantum Valley Ideas Laboratories, 485 Wes Graham Way, Waterloo, ON N2L 6R2, Canada
[2]5. Physikalisches Universität Stuttgart, Pfaffenwaldring 57, 70569 Stuttgart, Germany

## I. Detailed description of density matrix model

The three-photon system can be described using the time-dependent master equation for a 5-level ladder system, which is solved for the density matrix $\rho$ in steady state:

$$\dot{\rho} = \frac{i}{\hbar}[\rho, H] + \mathcal{L}(\rho), \tag{S1}$$

$H$ is the Hamiltonian of the system in the interaction picture, $\hbar$ is the reduced Planck's constant, and $\mathcal{L}$ is the Lindblad operator which describes decays and dephasing. We describe the Hamiltonian of this system in the interaction picture by:

$$H = \hbar \begin{pmatrix} 0 & \frac{\Omega_{895}}{2} & 0 & 0 & 0 \\ \frac{\Omega_{895}}{2} & -\Delta_2 & \frac{\Omega_{636}}{2} & 0 & 0 \\ 0 & \frac{\Omega_{636}}{2} & -\Delta_3 & \frac{\Omega_{2262}}{2} & 0 \\ 0 & 0 & \frac{\Omega_{2262}}{2} & -\Delta_4 & \frac{\Omega_{RF}}{2} \\ 0 & 0 & 0 & \frac{\Omega_{RF}}{2} & -\Delta_5 \end{pmatrix} \tag{S2}$$

The Hamiltonian includes terms involving the Rabi frequencies $\Omega_j = \mu_j E_j/\hbar$ of each state transition $j$, where the atomic dipole moments $\mu_j$ are averaged over the hyperfine states accessed in the setup. The detunings of each state are given by:

$$\Delta_2 = \Delta_{895} + k_{895} v \tag{S3a}$$

$$\Delta_3 = \Delta_2 + \Delta_{636} - k_{636} v \tag{S3b}$$

$$\Delta_4 = \Delta_3 + \Delta_{2262} + k_{2262} v \tag{S3c}$$

$$\Delta_5 = \Delta_4 + \Delta_{RF} \tag{S3d}$$

Where $v$ is the magnitude of the atom's velocity parallel to the laser beams and $k$ is the photon wave vector. For all simulations, the 895 nm and 636 nm lasers are held on resonance ($\Delta_{895} = \Delta_{636} = 0$), as is the radio frequency (RF) field ($\Delta_{RF} = 0$). A Doppler averaged value of element $\rho_{21}$ of the density matrix is obtained by solving Eqn. (S1) at individual velocities $v$, then weighted by the Boltzmann distribution and numerically integrated according to:

$$\rho_{21} = \int \sqrt{\frac{m}{2\pi k_B T}} \exp\left(-\frac{mv^2}{2k_B T}\right) \rho_{21}(v)\, dv \tag{S4}$$

The density matrix element $\rho_{21}$ can be used to determine an absorption coefficient $\alpha$, described by:

$$\alpha = \frac{N\mu_{895}^2 k_{895}}{2\epsilon_0 \hbar \Omega_{895}} Im(\rho_{21}) := \alpha_0 Im(\rho_{21}), \tag{S5}$$

where $N = 3 \times 10^{16}$ m$^{-3}$ is the atom number density at room temperature and $\epsilon_0$ is the permittivity of free space. We define the leading coefficient as $\alpha_0$. The total transmitted intensity $I$ of the probe laser through the vapour cell is described by the Lambert-Beer law using the absorption coefficient in Eqn. (S5), the length of the vapour cell $L$, and the incident probe intensity $I_0$:

$$I = I_0 \exp(-\alpha L) \tag{S6}$$

## II. Collisions and Pairwise Interation Potentials

To evaluate the possible role of collisional broadening on the three-photon EIA linewidth, we calculate pair-wise potentials for the $42P_{3/2}$ Rydberg pair states shown in Figure S1. These calculations include both dipole and quadrupole interactions [1,2]. The $42P_{3/2} - 42P_{3/2}$ interaction is attractive, and we expect atoms to experience a detuning comparable to or larger than the laser linewidths ($\geq$20 kHz) when they are within $R_{nn} \sim 4.4$ μm of one another based on Figure S1b. Given an average thermal velocity $\bar{v}$ = 200 m/s and an atomic density $N = 3 \times 10^{16}$ m$^{-3}$, we expect this to occur at an infrequent rate of $\Gamma_{coll} \sim \rho_{44} N \pi R_{nn}^2 \bar{v} \sim 2\pi \times 2$ kHz, which produces negligible collisional dephasing and broadening.

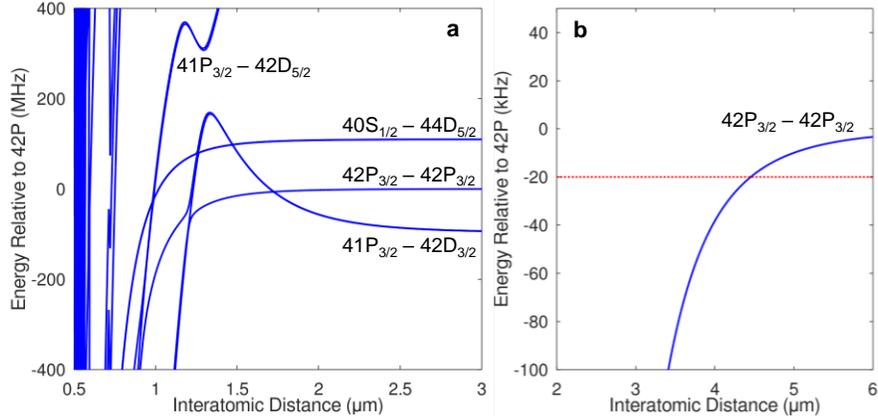

Fig. S1. Pairwise interaction potentials calculated for the $42P_{3/2}$ and nearby Rydberg states. (a) A state-changing quadrupole-allowed collision is possible at an interatomic distance of ~1.2 μm, but is very unlikely due to the steep avoided crossing and extremely low Landau-Zener curve crossing probability. (b) Vertically magnified plot of (a) focused on small frequency shifts. For a $42P_{3/2} - 42P_{3/2}$ interaction to result in detuning comparable to laser linewidths ~20 kHz (red dashed line), the atoms must be closer than ~4.4 μm, which occurs infrequently.

## III. Optimal Conditions for RF Sensitivity

In Figure S2 we vary the optical powers, and therefore the Rabi frequencies, of the three lasers to characterize the optimal conditions for EIA and RF sensitivity. Figure S2a shows the measured amplitude of the three-photon EIA peak with no RF field applied, obtained by comparing the probe laser's photodiode voltage with a resonant and a far-detuned 2262 nm laser. Figures S2b-c show the change in resonant absorption that occurs in response to an RF electric field (E-field) of 1370 μV/cm. A larger three-photon EIA peak provides a better signal-to-noise ratio in the Autler-Townes regime, while a larger response to RF E-fields is better for detection in the amplitude regime.

RF sensitivity increases with both $\Omega_{2262}$ and $\Omega_{636}$, but more rapidly with $\Omega_{2262}$. The trends are consistent with density matrix modeling shown in Figure S2d, which extrapolates to higher Rabi frequencies than possible in our current experimental setup. A transit time broadening of $2\pi \times 1$ MHz is used in the model, reflecting the smaller beam sizes used for our sensitivity measurements. The effects of collisions and ionization are not included, which will likely broaden the linewidth and reduce RF sensitivity at the upper right end of the map. The results indicate that both coupling Rabi frequencies should be increased to improve RF sensitivity.

The experiment exhibits an optimal value of $\Omega_{895}$ for maximum RF sensitivity, which occurs at about half the $\Omega_{895}$ that achieves a maximum 3-photon EIA peak height. The optimal value for $\Omega_{895}$ does not appear to change significantly as $\Omega_{636}$ and $\Omega_{2262}$ are varied.

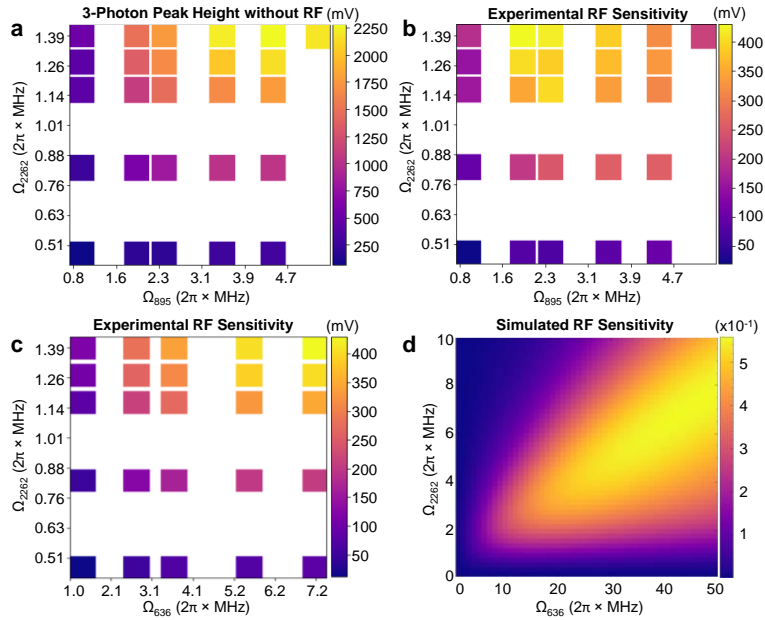

Fig. S2. Optimal Rabi frequencies for highest RF sensitivity. (a) Height of the measured 3-photon absorption peak without RF applied at $\Omega_{636} = 2\pi \times 7.2$ MHz. (b) Measured change in absorption with an incident RF E-field of ~1370 µV/cm and $\Omega_{636} = 2\pi \times 7.2$ MHz. (c) Measured change in absorption with an incident RF E-field of ~1370 µV/cm and $\Omega_{895} = 2\pi \times 1.9$ MHz. (d) Simulated map of the percent change in the absorption coefficient α with and without a 196 µV/cm RF E-field present at a fixed $\Omega_{895} = 2\pi \times 1.9$ MHz.

We measure the RF E-field sensitivity at the best experimentally accessible Rabi frequencies of $\Omega_{2262} = 2\pi \times 2.4$ MHz, $\Omega_{636} = 2\pi \times 7.2$ MHz, and $\Omega_{895} = 2\pi \times 1.9$ MHz. This results in a linewidth of ~1.5 MHz, as shown in Figure S3. Figure S3 also shows the change in absorption induced by the RF E-field as a function of the 2262 nm laser detuning, measured using a 20 kHz square wave amplitude modulation on the RF E-field and demodulated with a lock-in amplifier. The effective linewidth of the absorption change is spectrally narrower, $\sim 2\pi \times 650$ kHz, than the EIA peak.

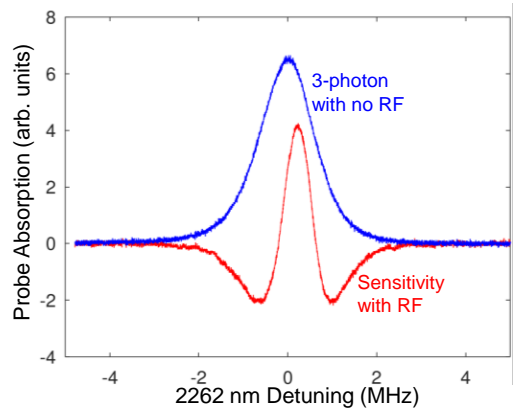

Fig. S3. Probe absorption with no RF E-field applied (blue), under laser conditions optimized for RF E-field sensitivity. The signal is measured using a lock-in amplifier with the 2262 nm laser amplitude modulated at 60 kHz. Sensitivity to an RF E-field of 1134 µV/cm is shown in red, measured using a lock-in amplifier with the RF amplitude modulated using a square wave at 20 kHz instead of modulating the 2262 nm laser.

## IV. Pulse Shapes

In Figure S4 we show experimental atomic responses of the three-photon system to 3724 µV/cm RF E-field pulses of various pulse durations, measured at room temperature. These have been averaged over $10^5$ cycles to reduce noise, and we use these as the templates required for matched filtering. Shorter pulses contain less total energy and produce a smaller matched filter peak along with larger noise due to higher bandwidth. Responses exhibit ~0.5 µs rise and fall times, influenced by the transit time of atoms through the beams. Due to this finite atomic response time, pulses below 0.5 μs durations do not reach the full steady-state height and therefore produce a worse signal-to-noise ratio for detection.

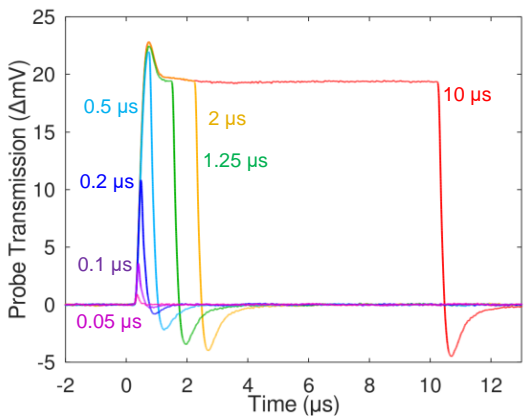

Fig. S4. Atomic response to 3724 μV/cm RF pulses of different durations, averaged over $10^5$ cycles.

In Figure S5, we characterize pulse shape with respect to different RF E-field strengths. Larger RF E-field amplitudes produce taller pulses and also produce larger overshoots on both edges of the pulse. The overshoot on the leading edge of the pulse also becomes slightly steeper (i.e., it has a faster response) with larger RF E-field amplitudes.

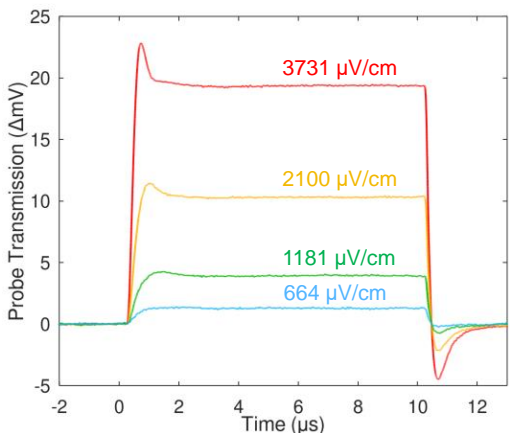

Fig. S5. Experimental atomic response to 10 μs long RF pulses of different field strengths, averaged over $10^5$ cycles. The overshoots on the leading and trailing edges are more pronounced and slightly steeper at stronger RF E-fields.

In practice, we do not observe any significant difference in the matched filter's output amplitude between using a pulse shape template that is matched to the incident RF field strength compared to one at an incorrect field strength. Figure S6 shows the results of processing 10 μs RF pulses using two different pulse templates. The templates have been normalized to have the same pulse height, but have different amounts of overshoot on the leading and trailing edges. There is no clear distinction between the results from the two templates at any RF field strength.

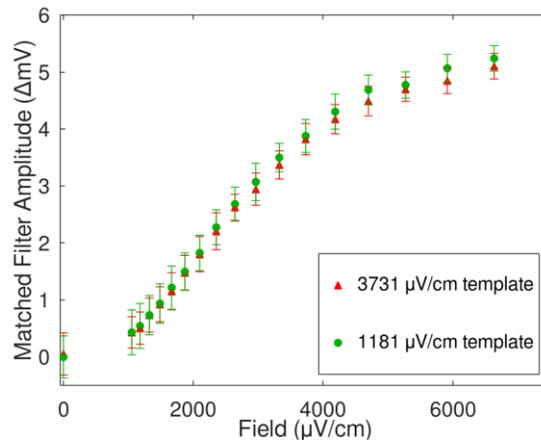

Fig. S6. Sensitivity to 10 µs RF pulses after passing through a matched filter constructed using either a 3731 µV/cm template (red) or the 1181 µV/cm template (green) from Fig. S5.